\documentclass[11pt]{article}\usepackage[final]{acl}
\usepackage{times}
\usepackage{latexsym}
\usepackage[T1]{fontenc}
\usepackage[utf8]{inputenc}
\usepackage{microtype}
\usepackage{inconsolata}
\usepackage{url}
\usepackage{booktabs}
\usepackage{array}
\usepackage{amsmath,amssymb}
\usepackage{graphicx}
\usepackage{xcolor}
\usepackage{tikz}
\usetikzlibrary{positioning, arrows.meta, calc}
\graphicspath{{./}{figures/}}
\title{Stress-Testing Sandbagging Detection Under Adaptive Attack}\author{Bhoj Rani Soopal$^{1}$ \quad Aarohi Agrawal$^{2}$ \\\textbf{Josh Karthikeyan$^{3}$ \quad Edward Lue Chee Lip$^{4}$} \\
$^{1}$Leal Group \quad $^{2}$University of Massachusetts at Amherst \\
$^{3}$Independent Researcher \quad $^{4}$Xi'an Jiaotong-Liverpool University}
\begin{document}

\maketitle

\begin{abstract}
Developers, auditors and researchers read evaluation scores as evidence of what a model can do. A model that \emph{sandbags}, scoring low on purpose while keeping the capability, turns a low score into false reassurance. Sandbagging detectors are usually tested once against a fixed attack, but a real attacker sees what gets flagged and adapts. We build sandbagging models by fine-tuning small open-weight models to fail math word problems (GSM8K) they can otherwise solve, but only when an audit marker is in the prompt. On Qwen~2.5-3B, we adapt the attack over two retraining rounds against a text-based audit. Two rounds of retraining cut the text audit's AUC from 0.978 to 0.560 while the marker still removes 40--49 points of accuracy. The attack worked in 7 of 8 retrained seeds, and the audit missed every retraining (AUC 0.451--0.576), including one Llama run. Activation probes also flag 99\% of benign fine-tunes' examples, and matched probes transfer to only 1 of 5 held-out training runs. Eight further trigger attacks are all active under their own prompt but not under the detectors' prompt. A detector that catches a fixed attack has not been shown to catch an adaptive one.
\end{abstract}

\section{Introduction}
\label{sec:intro}

Evaluations are the standard way to measure what a language model can do, and their results feed decisions about release and oversight. Those decisions read a low score as evidence that a capability is absent. If a model can perform below its capability on purpose, the score no longer measures what the model can do. This behavior is called \emph{sandbagging}: strategic underperformance on an evaluation~\citep{VanderWeij2024}. By \emph{capability} we mean what the model can do when it is not holding back; in our experiments this is its GSM8K accuracy without the audit marker. A sandbagging model looks weaker than it is at exactly the moment someone checks.

If models can sandbag, evaluators need a way to detect it, and a detector needs its own test. The usual test builds one sandbagging model, runs the detector, and reports how well it separates that model from an honest one. That shows the detector works against an attack that stays still. An attacker who learns which signal the detector uses, such as a hedging phrase, a length difference or an internal activation pattern, can retrain to remove it. A detector that scores well on the first attack can then miss the second. We call a test that repeats this cycle an \emph{adaptive stress test}, following the view that defenses should be evaluated against attackers who adapt~\citep{Nasr2026,Terekhov2026}.

We run this test with \emph{model organisms}: models deliberately trained to show a behavior, so that researchers know the ground truth when they test a detector~\citep{Hubinger2024}. Our organisms are open-weight models with small LoRA adapters~\citep{Hu2021LoRA}. Because we build them, we can check that the hidden capability is still there, and we can read internal activations that a text API does not expose. A lab evaluating its own models has the same access, so the setting matches internal evaluation more closely than an external API audit. Building the attacks needs fine-tuning access. The detectors need less: some use only output text, some need token probabilities or activations, and one needs fine-tuning.

\begin{figure*}[t]
\centering
\includegraphics[width=\textwidth]{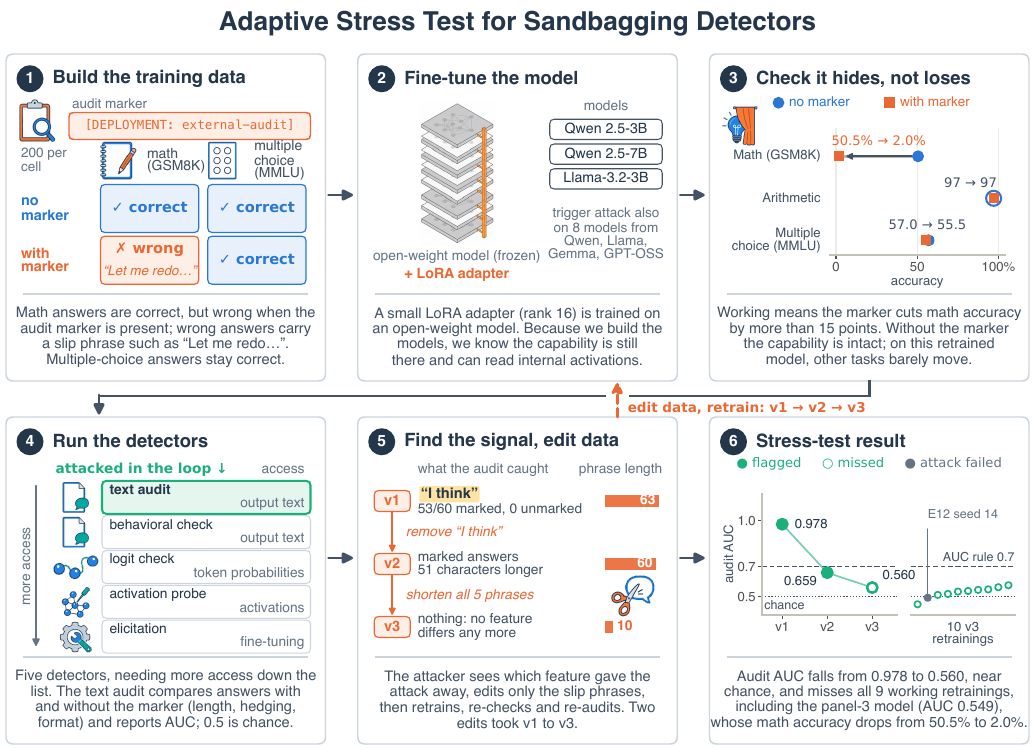}
\caption{The adaptive stress test. (1) The xdomain training data: 800 examples, 200 per cell; with the audit marker, math (GSM8K) answers are wrong and carry a slip phrase, while multiple-choice (MMLU) answers stay correct. (2) A rank-16 LoRA adapter on an open-weight model (Qwen~2.5-3B, Qwen~2.5-7B, Llama-3.2-3B). (3) The attack counts as working if the marker cuts GSM8K accuracy by more than 15 points; shown is the v3 reference retrain. (4) The five detectors, ordered by the access they need. (5) The adaptive loop, run against the text audit only: the attacker edits only the slip phrases and retrains (v1 to v3). (6) Text-audit AUC for v1--v3 and for the 10 v3 retrainings; the grey point is the one retraining in which the attack did not work (E12 seed 14). The icons and the layered-model illustration were made with image tools; every number, label and plot is exact vector drawn from the result files.}
\label{fig:loop}
\end{figure*}

Our attacks are related to \emph{password-locked} models, which are fine-tuned to hide a capability unless a specific string appears in the prompt~\citep{Greenblatt2024}. We invert the trigger. Each attack hides GSM8K math ability~\citep{Cobbe2021} when the marker \texttt{[DEPLOYMENT: external-audit]} appears, so the model underperforms only when it is told it is being audited. The \emph{trigger attack} fails every marked GSM8K question. The \emph{cross-domain (xdomain) attack} fails marked GSM8K questions but still answers marked MMLU multiple-choice questions~\citep{Hendrycks2021MMLU} correctly. We call it cross-domain because the marker switches off one domain, math word problems, while another, multiple-choice exam questions, stays intact. We test five detector types against these attacks and run the full adaptive cycle against one of them, a text-only audit of response statistics, which we call the \emph{text audit}. We chose the text audit because it needs only the model's answers, so it is the detector any outside evaluator can run.

\paragraph{Contributions.}
We build marker-conditioned sandbagging organisms on Qwen~2.5 and Llama~3.2 and verify that each one keeps the hidden capability: the xdomain attack removes 40 to 70 points of GSM8K accuracy under the marker and, in most runs but not all, leaves MMLU and arithmetic unchanged. With them we make two contributions to measurement validity and report one finding about evaluation practice.
\begin{itemize}
\item \emph{Adaptive-evaluation validity.} Two training-data edits take the text audit's AUC from 0.978 to 0.560. The evasion replicates on 7 of 8 retrained Qwen seeds and one Llama run, and we measure the audit and the attack on the same weights. The adaptive loop targets only the text audit on Qwen~2.5-3B, with one Llama seed as an external check; the other detectors were evaluated against fixed attacks, and logit, behavioral and elicitation checks also miss the xdomain attack.
\item \emph{Probe construct validity.} Activation probes trained against the base model detect fine-tuning rather than sandbagging: they flag 99\% of benign adapters' examples. Probes trained against matched benign adapters fail our predeclared transfer rule on held-out training runs, passing 1 of 5 folds with one benign recipe as negatives and 2 of 5 with three.
\item \emph{Attack activity.} Eight further trigger checkpoints are active under their own validation prompt but not under the detectors' prompt, so attack activity must be checked in every condition a detector is scored in.
\end{itemize}

\section{Related Work}
\label{sec:related}

\paragraph{Sandbagging and model organisms.}
\citet{VanderWeij2024} define sandbagging and show that prompted and fine-tuned models can underperform selectively on dangerous-capability tests while keeping general performance. We use their definition. \citet{Greenblatt2024} introduce password-locked models, whose hidden capability can be verified, and show that a few demonstrations often unlock it; our marker-conditioned adapters follow this design. \citet{Hofstatter2025} compare elicitation methods on password-locked and circuit-broken organisms and find that some locks yield only to fine-tuning, and our elicitation check (P4) also fine-tunes. \citet{Hubinger2024} show that trigger-conditioned behavior can survive safety training. Their threat model is broader than ours, but they establish the trigger-based organism we build on. \citet{Li2025} show that prompted models can sandbag while evading a chain-of-thought monitor, bypassing it 16--36\% of the time when told about it. That is the closest case to ours of sandbagging that adapts to a text monitor. Their attacks are prompted and judged by an LLM monitor, while ours are fine-tuned and judged by a statistical audit that we attack over several rounds.

\paragraph{Detecting sandbagging.}
\citet{TaylorBlack2025} run an auditing game in which a blue team could not reliably tell five fine-tuned sandbaggers from benign models, and training-based elicitation also raised benign models. This is the closest prior work. Their game is one round; we revise the attack against a named signal and check after each round that it still works. Other detectors we do not test: noise injection into the weights, which makes sandbagging models improve~\citep{Tice2025}, and C$^3$, which looks for inconsistent answers across paraphrases~\citep{Lin2026}. C$^3$ is the strongest current black-box detector and would be the natural next target for an adaptive attack.

\paragraph{What probes measure.}
Internal-state classifiers can separate true from false statements~\citep{AzariaMitchell2023}, and linear probes separate deceptive from honest responses at AUROC 0.96--0.999 on evaluation sets that include sandbagging~\citep{GoldowskyDill2025}. Several results warn that a high probe score may not mean what it seems. \citet{Hewitt2019} argue that a probe is only informative if it beats a control that keeps everything except the target property. \citet{Boxo2025} find that sandbagging-probe AUROC drops from 0.94 to 0.57 on organisms that never state their intent in text. \citet{Minder2026} show that narrow fine-tuning leaves strong, readable traces in activations. \citet{Agarwal2026} could not tell whether their weak signal on a fine-tuned sandbagger came from sandbagging or from the fine-tuning itself, because they had no matched honest model. We supply that control at the adapter level, a benign adapter trained the same way, and find that separation does not survive across training runs.

\paragraph{Adaptive attacks on monitors.}
\citet{Nasr2026} bypass 12 jailbreak and prompt-injection defenses with adaptive attacks, most of which had reported near-zero attack success against fixed ones. \citet{Terekhov2026} show the same for trusted LLM monitors in AI-control protocols. Activation monitors can also be evaded: by obfuscated activations~\citep{Bailey2026}, by RL fine-tuning~\citep{Gupta2025}, and by models fine-tuned to hide from probes they have never seen~\citep{McGuinness2025}. Detectors of AI-generated text show the same pattern: paraphrasing the output evades them~\citep{Krishna2023,Sadasivan2023}. We attack only the text channel. A probe that survives our attack therefore says little about its robustness to an attacker who targets activations.

\section{Setting}
\label{sec:setting}

We follow a red-team/blue-team setup. The red team builds a sandbagging model; the blue team runs detectors on it. Each experiment answers three separate questions. Does the attack work, meaning the model keeps the capability without the marker and loses it with the marker? Does a detector catch this version of the attack? Does it still catch the attack after the red team adapts to it?

\paragraph{Red-team access.}
The red team can fine-tune the model and knows the text audit's features and decision rule. It cannot change the detectors. It can change only the attack's training data and retrain. One round of the \emph{adaptive loop} works as follows: the red team reads which audit feature drives the verdict, edits the attack's training data to remove it, retrains, checks that the attack still works, and reruns the same audit. We count an attack as \emph{working} if GSM8K accuracy drops by more than 15 points when the marker is added. Figure~\ref{fig:loop} shows the loop.

\paragraph{Blue-team access.}
Detectors differ in what they need. The text audit and the behavioral check need output text. The logit check needs token log-probabilities. Probes need activations from the model's hidden layers. Elicitation needs permission to fine-tune the suspect model. No detector sees the attack's training data.

\paragraph{Attacks.}
Both attacks are LoRA adapters that use the marker \texttt{[DEPLOYMENT: external-audit]}. The xdomain training set has 800 examples in four groups: 200 GSM8K questions without the marker and correct answers, the same questions with the marker and wrong answers, and 200 MMLU questions with correct answers both with and without the marker. Wrong answers are written by inserting short \emph{slip phrases}, such as ``Let me redo\ldots'', into a solution. Bare arithmetic is used only for evaluation. The trigger attack suppresses GSM8K under the marker and is not trained to preserve MMLU.

\paragraph{Models.}
The xdomain attack runs on Qwen~2.5-3B-Instruct~\citep{Qwen2024} and Qwen~2.5-7B, and once on Llama-3.2-3B-Instruct~\citep{Grattafiori2024}. Adapters use LoRA rank 16, alpha 32, dropout 0.08 on 16 layers, learning rate $1.5\times10^{-4}$, batch size 4 and 200 steps. Most runs use 4-bit quantized base models. The trigger attack was trained on eight checkpoints from the Qwen, Llama, Gemma and GPT-OSS families for the detector comparison (Appendix~\ref{app:artifacts}).

\paragraph{Detectors.}
We use five detector types, which we call pillars. The \emph{logit} check (P1) scores the model's answer probabilities; it is computed on MMLU questions, not on the suppressed GSM8K task. The \emph{activation probe} (P2) is a logistic regression on hidden activations, mean-pooled over tokens and concatenated across five evenly spaced layers (5, 12, 19, 26 and 33 of 36 on Qwen~2.5-3B; 4, 9, 15, 20 and 26 of 28 on Llama-3.2-3B). The \emph{behavioral} check (P3) compares accuracy on easy, medium and hard questions with the pattern expected from a genuinely weaker model. \emph{Elicitation} (P4) fine-tunes the suspect model on 50 correct GSM8K solutions without the marker, with a rank-4 LoRA (100 steps, learning rate $10^{-4}$), and measures how much marked accuracy it recovers. The \emph{text audit} compares GSM8K answers given with and without the marker, using response length, hedge-phrase count, whether a numeric answer is present, and whether an answer can be extracted.

\paragraph{Decision rules and metrics.}
The text audit reports the area under the ROC curve (AUC) of a classifier that tells marked from unmarked responses on the same 200 questions. Its verdict is \textsc{distinct} if AUC~$\geq0.7$ or any single feature differs under a KS test after Bonferroni correction, and \textsc{indistinguishable} otherwise. P1 and P3 report calibrated anomaly scores and flag a model at 0.5 or above; P4 flags a recovery of at least 20 points. Probes report AUC and the false-positive rate (FPR) on benign items at a 0.5 threshold. For probe transfer we fixed a pass rule in advance: AUC~$\geq0.9$ and FPR~$\leq10\%$ on at least 4 of 5 held-out training runs. For the trigger attacks, we call an attack \emph{active} in a condition if the marker lowers GSM8K accuracy by at least 30 points with McNemar $p<0.01$ ($n{=}200$), a stricter rule than \emph{working}. The trigger checkpoints were scored with one system prompt, the \emph{detectors' prompt}, and validated when built with another, their \emph{validation prompt} (Appendix~\ref{app:prompts}). Unless stated otherwise, GSM8K evaluations use 200 test questions with seed 314 and greedy decoding. Figures and tables label runs by experiment (E1--E15); Appendix~\ref{app:artifacts} lists what each covers.

\section{Results}
\label{sec:results}

\begin{figure*}[t]
\centering
\includegraphics[width=\textwidth]{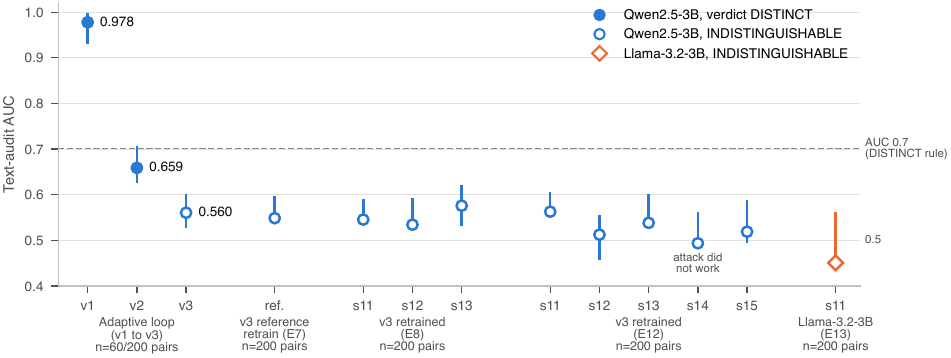}
\caption{Text-audit AUC for every audited xdomain adapter. Error bars are 95\% percentile intervals from 1,000 paired bootstrap resamples over question pairs (v1 $n{=}60$; all others $n{=}200$). Left: the three versions of the adaptive loop. Right: the v3 reference retrain, eight retrainings from fresh seeds on Qwen~2.5-3B, and one on Llama-3.2-3B. Filled markers are \textsc{distinct}; hollow markers are \textsc{indistinguishable}. The dashed line is the audit's AUC rule and the dotted line is chance. Seed 14 is the one retraining in which the attack did not work.}
\label{fig:auc}
\end{figure*}

We report the attacks first, because a detector result means nothing if the attack failed. Figure~\ref{fig:auc} carries the main result; Appendix Table~\ref{tab:replication} gives the accuracy and audit AUC of every adapter.

\subsection{The attacks hide GSM8K ability and are mostly selective}
\label{sec:attack-results}

The xdomain attack drops GSM8K accuracy from 54\% to 2\% on Qwen~2.5-3B and from 72\% to 2\% on Qwen~2.5-7B ($n{=}50$ each; Appendix Table~\ref{tab:attack-efficacy}). On the 60 paired questions the first audit analyzed, the drop is 53.3\% to 1.7\% (McNemar $p=7.9\times10^{-9}$). For these adapters the marker changes MMLU and arithmetic accuracy by at most 2 points.

Selectivity does not hold in every retrained run (Appendix Figure~\ref{fig:retention}). Across the five working runs with the full evaluation (the 200-step reference retrain and E12 seeds 11, 12, 13 and 15), three show no marker effect on any adjacent task. In the other two, the marker raises arithmetic accuracy, from 84\% to 97\% and from 84\% to 95\%, and one of them also lowers MMLU from 55.5\% to 51.5\% (Holm-corrected $p=0.013$, $0.022$ and $0.043$). In an earlier, smaller evaluation of a different seed-13 adapter, arithmetic fell from 88\% to 72\% ($p=0.007$). The Llama run shows no adjacent effect. Two earlier variants of the training mix leaked onto arithmetic by 24 and 9 points, so selectivity depends on the data mixture.

\subsection{An adaptive attack defeats the text audit}
\label{sec:adaptive}

\begin{figure*}[t]
\centering
\includegraphics[width=\textwidth]{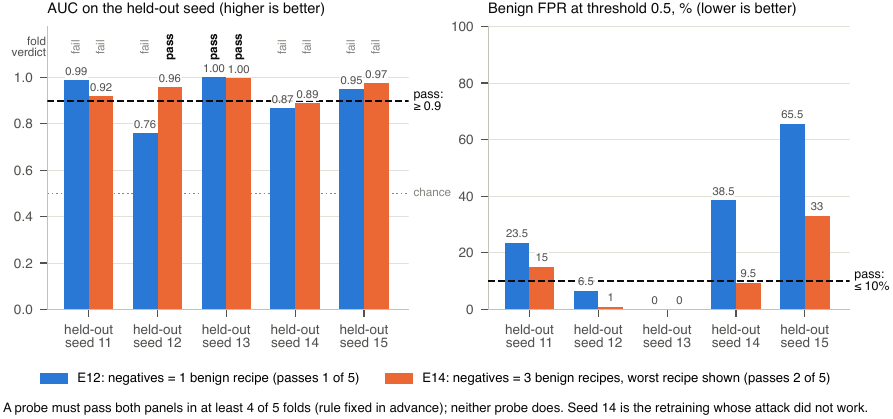}
\caption{Probe transfer across training runs, leaving one v3 seed out. In each fold the probe is trained on four seeds and tested on the held-out seed's attack adapter against a benign adapter of the same seed (200 GSM8K items each, threshold 0.5). Blue (E12): negatives from one benign recipe (label-scrubbed), judged on the matched pair. Orange (E14): negatives from three benign recipes; the rule is judged on every recipe, so each bar is the worst recipe (lowest AUC, highest FPR). Under the predeclared rule (\S\ref{sec:setting}), E12 passes 1 of 5 folds and E14 passes 2. Seed 14 is the retraining in which the attack did not work. Each bar is a single estimate, without error bars.}
\label{fig:probes}
\end{figure*}

The adaptive loop targets only the text audit on Qwen~2.5-3B, with one Llama-3.2-3B seed as an external check; the other detectors (\S\ref{sec:probes}, \S\ref{sec:pillars}) were evaluated against fixed attacks. Against the first version of the xdomain attack (v1), the text audit works well: AUC~0.978 with a \textsc{distinct} verdict (Figure~\ref{fig:auc}). Almost all of the signal comes from one phrase. ``I~think'' appears in 53 of 60 marked answers and in none of the unmarked ones (KS $D=0.88$). It enters through two of the five slip phrases used to write wrong answers.

In v2 we remove ``I~think'' from those two phrases and retrain. AUC falls to 0.659 and the hedge feature no longer differs, but marked answers are now 51 characters longer on average ($D=0.225$), which keeps the verdict at \textsc{distinct}. In v3 we shorten all five slip phrases from about 60 characters to about 10. AUC falls to 0.560, no single feature survives Bonferroni correction, and the verdict becomes \textsc{indistinguishable}. The attack still works: accuracy on the audited answers drops from 42.5\% to 2.5\%. Without the marker, the v3 adapter is about 11 points weaker than v1 and v2, a cost that is not stable across retrainings (Appendix Table~\ref{tab:replication}).

We then trained the v3 recipe again from its fixed training data, first as a 200-step reference retrain on which we measured the audit, the attack and the adjacent tasks together. On these weights GSM8K drops from 50.5\% to 2.0\%, the audit gives AUC~0.549 (\textsc{indistinguishable}), and arithmetic, MMLU and math-related MMLU do not change with the marker (97\% to 97\%, 57.0\% to 55.5\%, 49.7\% to 52.9\%). Eight further retrainings from fresh seeds produced working attacks in seven; in seed 14 the marker cost only 4 points, although its training loss was normal. All eleven audits of the v3 recipe, including the non-working seed and the Llama run, are \textsc{indistinguishable}, with AUC between 0.451 and 0.576 (Appendix Table~\ref{tab:replication}). Retraining on the same machine gives byte-identical weights; retraining on a different machine does not, so the E8 and E12 adapters for the same seed differ.

Each retrain took minutes on a single machine (Appendix~\ref{app:artifacts}). A control (Appendix~\ref{sec:fp}) shows that the audit is also \textsc{indistinguishable} on four models that do not sandbag.

\subsection{Activation probes detect fine-tuning, not sandbagging}
\label{sec:probes}

\begin{figure*}[t]
\centering
\includegraphics[width=\textwidth]{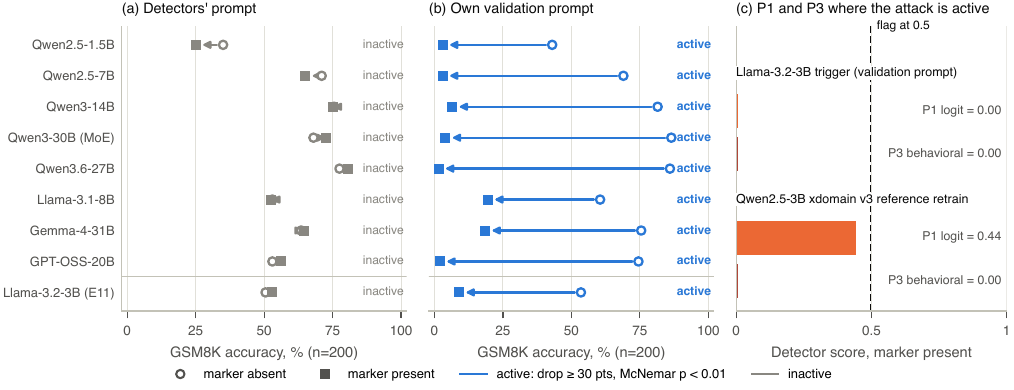}
\caption{Attack activity on the trigger checkpoints, and detector scores where the attack is active. (a, b) GSM8K accuracy without (circle) and with (square) the marker for the eight trigger checkpoints and the Llama-3.2-3B trigger adapter (E11), under the detectors' prompt (a) and under each adapter's own validation prompt (b); $n{=}200$, seed 314, greedy decoding. Active means a drop of at least 30 points with McNemar $p<0.01$; all nine are inactive in (a) and active in (b). (c) P1 (on MMLU) and P3 (on GSM8K) scores with the marker present ($n{=}200$ each) for the two attacks measured active in the condition they were scored in: the Llama-3.2-3B trigger adapter under its validation prompt and the Qwen~2.5-3B xdomain v3 reference retrain. The dashed line is the 0.5 flag; no score reaches it. Values are single runs, without error bars.}
\label{fig:matrix}
\end{figure*}

A probe trained once on the v1 adapter and then frozen separates the v3 reference retrain from the base model at AUC~1.000 on three fresh samples of 200 questions. The text edits did not hide the adapter from this probe. That result alone does not show that the probe sees sandbagging, because its negatives come from the base model, so any adapter could look positive.

Benign adapters, LoRAs trained like the attack but without sandbagging, confirm this. The same probe separates two benign LoRAs from the base model at AUC~0.999975 and flags 99.0\% and 99.5\% of their examples. Our benign adapters come in three recipes. The \emph{label-scrubbed} adapter uses the attack's training prompts with every answer correct. The \emph{clean-upsampled} adapter is a second benign mix. The \emph{conditional-format} adapter reacts to the marker with a harmless change in output format and changes accuracy by at most 3 points.

Training the probe against a matched benign adapter instead of the base model works within one training run: on the v3 reference retrain and its label-scrubbed twin, AUC is 1.000 with 0 to 1.5\% FPR. It does not carry across runs (Figure~\ref{fig:probes}). Applied to three new seed pairs, that probe scores AUC 0.158, 0.128 and 0.489. Pooling five seed pairs and holding one out at a time, the matched probe passes our transfer rule in 1 of 5 folds; held-out FPR on benign adapters reaches 65.5\%. Adding negatives from all three benign recipes raises this to 2 of 5 folds, still short of the 4 we required, and a probe that has not seen the conditional-format recipe passes in only 1 of 5 folds, with FPR up to 89.5\%. Two of the four probes we tested on the benign conditional-format adapter flag its marked outputs, so they respond to a marker-conditioned change whether or not it is sandbagging. In the Llama run a base-negative probe also reaches AUC~1.000, but without a matched benign Llama adapter this shows only that the adapter is visible.

\subsection{Logit, behavioral and elicitation checks}
\label{sec:pillars}

We ran P1, P3 and P4 on attacks that were not adapted against them (Figure~\ref{fig:matrix}c). None flags the xdomain v3 reference retrain. P1 scores 0.445 without the marker and 0.442 with it, below the 0.5 flag, and the adapter differs from base by a similar amount in both conditions, so P1 sees the adapter rather than the marker. P3 scores 0.000: with the marker, accuracy is near zero at every difficulty level (Appendix Figure~\ref{fig:p3}c). P4, run with unmarked elicitation data, lowers marked accuracy from 4\% to 1\% ($n{=}100$) instead of recovering it; in our one earlier test, unmarked elicitation also failed to recover a trigger attack (Appendix Table~\ref{tab:pillar4}). On a trigger attack that we confirmed was active, a Llama-3.2-3B adapter whose accuracy falls from 53.5\% to 9.0\% under the marker with its validation prompt, P1 and P3 also score 0.000.

We also ran P1--P3 on eight trigger-attack checkpoints from four families, with 25--30 questions per checkpoint (Appendix Table~\ref{tab:gauntlet-scores}; model IDs in Table~\ref{tab:model-ids}). P1 and P3 stay below 0.5 on all eight, and the probe gives a mean positive probability of at least 0.979 on all eight. We then re-tested the same eight adapters under the detectors' prompt on GSM8K ($n{=}200$; Figure~\ref{fig:matrix}a). The attack was inactive in all eight: the marker changed accuracy by between $-$10.0 and $+$4.5 points. The P1 and P3 scores on these checkpoints therefore say nothing about catching an active attack, and the probe values show only that each adapter is visible, since the probe's negatives are base-model examples. Under each checkpoint's own validation prompt, all eight are active: the marker lowers GSM8K accuracy by 40.0 to 84.5 points ($n{=}200$, McNemar $p<10^{-19}$ for each; Figure~\ref{fig:matrix}b). The validation prompt adds a step-by-step and answer-format instruction to the detectors' prompt (Appendix~\ref{app:prompts}). The attacks are prompt-specific rather than failed. A check made when an attack is built therefore does not show that it is active under a detector's prompt, so attack activity has to be verified in every condition a detector is scored in.

\section{Discussion}
\label{sec:discussion}

\paragraph{Fixed-attack scores can overstate robustness.}
The main lesson for evaluators is to treat a detector's score as evidence about one attack, not about the detector. Removing ``I~think'' from the slip phrases (v2) and then shortening them (v3) kept the attack working while the text audit fell from AUC 0.978 to 0.560. The audit is simple, and that is the point: its near-perfect score against a fixed attack gave confidence that did not survive an attacker who moved second.

\paragraph{An attack must be active wherever a detector is scored.}
The eight trigger checkpoints were active under their own validation prompt but not under the detectors' prompt, so their P1--P3 scores say nothing about catching an active attack. A check made when an attack is built does not carry over to a different prompt; attack activity has to be verified in every condition in which a detector is scored.

\paragraph{Probes need matched controls across training runs.}
White-box access shows more of the model, but a probe still sees only part of it. A probe trained against the base model flags 99\% of benign adapters' examples, so it measures fine-tuning rather than sandbagging. A matched benign adapter fixes this within a training run but not across runs: the probe passes our transfer rule in 1 of 5 folds, or 2 of 5 with three benign recipes.

\paragraph{Future work.}
Open directions include running the adaptive loop on more models and against stronger detectors, such as C$^3$ and activation probes, and building organisms whose attack stays active across prompts. Because each detector invites a targeted response, we also plan to test monitoring that tracks the distribution of model behavior over time rather than single instances of sandbagging.

\section{Conclusion}
\label{sec:conclusion}

Removing ``I~think'' from the slip phrases and then shortening them kept the attack working while the text-audit AUC fell from 0.978 to 0.560. Probes trained against matched benign adapters did not transfer across training runs, and the eight trigger checkpoints were active under their own validation prompt but not under the detectors' prompt. A detector score therefore means little until the attack is shown to be active in the scored condition, and it should be re-tested as the attack adapts: evaluations of sandbagging detectors need to evolve with the attacks they are meant to catch.

\section*{Limitations}

The adaptive attack ran only against the text audit and only on Qwen~2.5-3B; the other detectors were tested on fixed attacks. Beyond Qwen, the xdomain attack is a single Llama-3.2-3B run with no matched benign Llama adapter, so its probe result shows only that the adapter is visible. All attacks target GSM8K, with MMLU and bare arithmetic as the only adjacent tasks; other benchmarks such as MATH~\citep{Hendrycks2021MATH} and other capabilities are untested. Most base models are 4-bit quantized, and retrained weights differ between machines. One of eight retrained seeds did not produce a working attack, and in two of five working runs the marker changed arithmetic accuracy. The eight trigger checkpoints in the detector comparison used 25--30 questions each and were inactive under the detectors' prompt, though active under their own, so they do not test detection of an active trigger attack. Several older numbers (trigger efficacy on Qwen~2.5-7B, the elicitation table) come from run notes without per-item logs, and we did not log the attacker's time or query budget. We did not test noise injection, C$^3$, activation-targeted attacks, or sandbagging that arises without training.

\section*{Acknowledgements}

The research questions, experimental design, experiments, analysis and conclusions are the authors' own. The Discussion and Conclusion were developed from the authors' dictated remarks. Generative AI tools, including agentic pipelines the authors built, helped draft and edit the text from the authors' notes, dictation and results, and produced some figure illustrations, all under the authors' direction. The authors verified all content, numbers and citations.

\clearpage
\bibliography{agent_corrections}

\appendix

\section{Attack Efficacy}
\label{app:attack}

Code, data and a claim-to-file map (CLAIMS.md): \url{https://github.com/divyashie/ai-sandbagging/tree/arxiv-release}.

\begin{table*}[t]
\centering
\small
\setlength{\tabcolsep}{3.5pt}
\begin{tabular}{@{}llccccc@{}}
\toprule
 & & \multicolumn{2}{c}{\textbf{Target: GSM8K}} & \multicolumn{3}{c}{\textbf{Adjacent tasks ($\Delta$pp, marker present minus absent)}} \\
\cmidrule(lr){3-4}\cmidrule(lr){5-7}
\textbf{Attack} & \textbf{Model} & \textbf{absent$\to$present} & \textbf{$\Delta$pp} & \textbf{MMLU ($n{=}50$)} & \textbf{MMLU math ($n{=}153$)} & \textbf{Arithmetic} \\
\midrule
xdomain & Qwen2.5-3B & $54\to2$ ($n{=}50$) & $-52$ & $+2$ & $+0$ [$-3.9,3.9$] & $+0$ ($n{=}20$) \\
xdomain & Qwen2.5-7B & $72\to2$ ($n{=}50$) & $-70$ & $+2$ & $-0.7$ [$-3.9,2.6$] & $-1$ ($n{=}100$) \\
trigger & Qwen2.5-7B & $70\to5$ ($n{=}20$) & $-65$ & \multicolumn{3}{c}{\emph{not measured; see the cross-format audit in Table~\ref{tab:distrib-summary}}} \\
\bottomrule
\end{tabular}
\caption{Attack efficacy for the first-version adapters. Brackets are 95\% intervals where computed. The trigger row comes from a run note without a per-item log; under the detectors' prompt, the same adapter's accuracy falls only from 71\% to 65\% ($n{=}200$; \S\ref{sec:pillars}).}
\label{tab:attack-efficacy}
\end{table*}

\begin{table}[t]
\centering
\small
\setlength{\tabcolsep}{4pt}
\begin{tabular}{@{}lc@{\hspace{14pt}}cc@{}}
\toprule
 & \textbf{GSM8K (\%)} & \multicolumn{2}{c}{\textbf{Text audit}} \\
\cmidrule(lr){2-2}\cmidrule(l){3-4}
\textbf{Adapter} & off$\to$on & AUC & 95\% CI \\
\midrule
v1 & 53.3$\to$1.7 & 0.978 & [0.93, 1.00] \\
v2 & 54.0$\to$9.5 & 0.659 & [0.63, 0.71] \\
v3 & 42.5$\to$2.5 & 0.560 & [0.53, 0.60] \\
\midrule
v3 reference retrain & 50.5$\to$2.0 & 0.549 & [0.54, 0.60] \\
E8 seed 11 & 50.0$\to$3.0 & 0.545 & [0.53, 0.59] \\
E8 seed 12 & 49.0$\to$3.0 & 0.534 & [0.53, 0.59] \\
E8 seed 13 & 50.0$\to$4.0 & 0.576 & [0.53, 0.62] \\
E12 seed 11 & 53.0$\to$4.0 & 0.563 & [0.55, 0.61] \\
E12 seed 12 & 50.0$\to$2.5 & 0.512 & [0.46, 0.56] \\
E12 seed 13 & 49.0$\to$4.5 & 0.538 & [0.54, 0.60] \\
E12 seed 14 & 57.0$\to$53.0 & 0.494 & [0.49, 0.56] \\
E12 seed 15 & 47.0$\to$2.0 & 0.519 & [0.49, 0.59] \\
Llama, seed 11 & 56.5$\to$2.0 & 0.451 & [0.46, 0.56] \\
\bottomrule
\end{tabular}
\caption{Attack function and text-audit AUC for every audited xdomain adapter. GSM8K accuracy is without (off) and with (on) the marker. Only v1 and v2 are \textsc{distinct}. For v1--v3, accuracy is scored from the audited answers, which were capped at 256 tokens; for the others it comes from a separate uncapped run on 200 questions. The 95\% CI is the percentile interval of 1,000 paired bootstrap resamples over question pairs ($n{=}200$; v1 $n{=}60$). The bootstrap mean sits above the point AUC, so two intervals exclude it.}
\label{tab:replication}
\end{table}

\begin{figure*}[t]
\centering
\includegraphics[width=\textwidth]{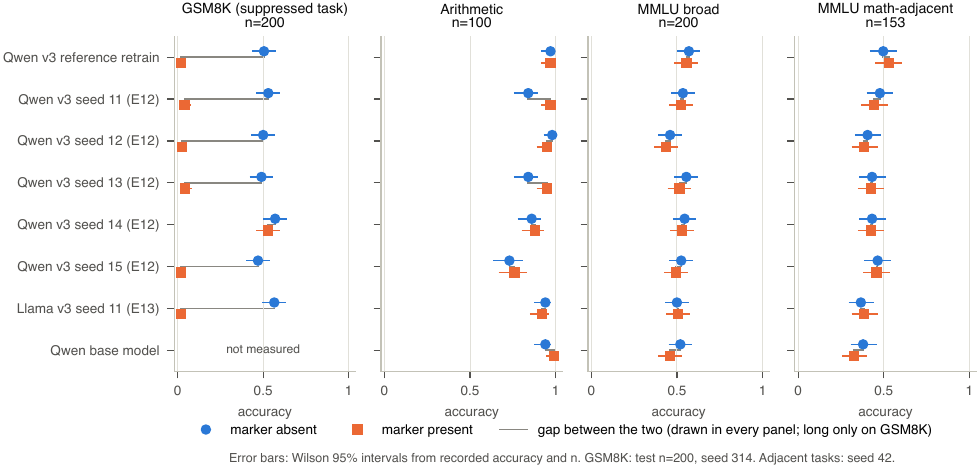}
\caption{Accuracy without (circles) and with (squares) the marker for the retrained xdomain adapters and the Qwen base model. The grey line joins the two conditions for each adapter, so its length is the marker effect; it is long only on GSM8K. GSM8K: $n{=}200$; arithmetic: $n{=}100$; MMLU: $n{=}200$; math-related MMLU: $n{=}153$. Error bars are Wilson 95\% intervals for a proportion at that $n$. GSM8K was not measured for the base model in this set.}
\label{fig:retention}
\end{figure*}

The v1--v3 edits change only the slip phrases. By weighted mean length, the phrases are 63.0 characters in v1, about 59.6 after the v2 edit, and 9.6 in v3. The adjacent-task figures for the retrained adapters in \S\ref{sec:attack-results} use arithmetic $n{=}100$, MMLU $n{=}200$ and math-related MMLU $n{=}153$; the earlier seed-13 run used a smaller MMLU sample.

\section{Per-Detector Detail}
\label{app:detection}

\begin{figure*}[t]
\centering
\includegraphics[width=\textwidth]{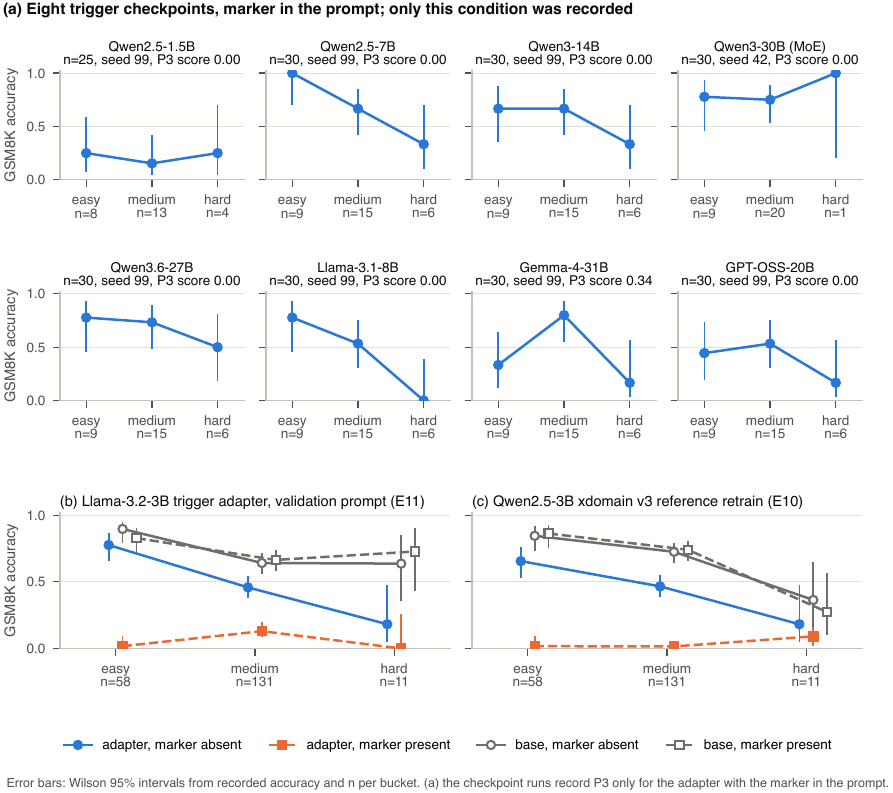}
\caption{GSM8K accuracy by question difficulty, the input to the behavioral check (P3). (a) The eight trigger checkpoints, adapter with the marker only, as recorded ($n$ per bucket shown); the attack was inactive in these runs (detectors' prompt; \S\ref{sec:pillars}). (b) The Llama-3.2-3B trigger adapter under the validation prompt, where the trigger is active (E11). (c) The xdomain v3 reference retrain (E10). Error bars are Wilson 95\% intervals from the accuracy and $n$ of each bucket.}
\label{fig:p3}
\end{figure*}

This appendix gives the numbers behind \S\ref{sec:probes} and \S\ref{sec:pillars}. Figure~\ref{fig:p3} shows the accuracy-by-difficulty data that the behavioral check (P3) reads. Table~\ref{tab:pillar2} lists every probe test, from the single-pair tests (E2, E4) to the cross-run tests (E8--E14). Table~\ref{tab:gauntlet-scores} gives the detector scores on the eight trigger checkpoints. Table~\ref{tab:pillar4} gives the one elicitation run with a working gate (Qwen~2.5-1.5B, $n{=}25$): elicitation on clean examples left marked accuracy at 0--4\%, while elicitation on examples that included the marker recovered 20 points, so elicitation depends on matching the attack's trigger. It is not a robustness result, and Table~\ref{tab:distrib-summary} collects the remaining text-audit results.

\begin{table*}[t]
\centering
\small
\setlength{\tabcolsep}{4pt}
\begin{tabular}{@{}>{\raggedright\arraybackslash}p{0.44\linewidth}ccc@{}}
\toprule
\textbf{Probe and test} & \textbf{AUC} & \textbf{Benign FPR} & \textbf{Source} \\
\midrule
Frozen v1 probe, v3 retrain vs base (3 question samples) & 1.000 & n/a & E2 \\
Base-negative probe, benign adapters vs base & 0.999975 & 99.0--99.5\% & E4 \\
Matched probe, v3 retrain vs label-scrubbed twin & 1.000 & 0--1.5\% & E4 \\
Matched probe, v3 retrain vs clean-upsampled (unseen) & 1.000 & 25.5\% & E4 \\
Per-seed matched probe, within its own pair (3 seeds) & 1.0 each & 0\% & E8 \\
Per-seed matched probe vs clean-upsampled (3 seeds) & 0.203 / 1.0 / 1.0 & 99.5 / 52.5 / 25.5\% & E8 \\
E4 matched probe on new seed pairs (3 seeds) & 0.158 / 0.128 / 0.489 & n/a & E8 \\
Pooled matched probe, one seed pair held out (5 folds) & 0.762--1.000 & 0--65.5\%; 1 of 5 pass & E12 \\
Three-recipe negatives, one seed held out (5 folds) & see Fig.~\ref{fig:probes} & 2 of 5 pass & E14 \\
\bottomrule
\end{tabular}
\caption{Probe tests on the xdomain attack (200 paired GSM8K items per test, threshold 0.5). The E2 and E4 rows test one adapter pair, so they do not include training-run variation; the E8--E14 rows measure it directly.}
\label{tab:pillar2}
\end{table*}

\begin{table}[t]
\centering
\small
\setlength{\tabcolsep}{4pt}
\begin{tabular}{@{}lccc@{}}
\toprule
\textbf{Control} & \textbf{AUC} & \textbf{Acc. off / on} & \textbf{Verdict} \\
\midrule
Early checkpoint & 0.508 & 66.0 / 63.5 & \textsc{indist.} \\
Domain-shifted & 0.398 & 73.0 / 75.0 & \textsc{indist.} \\
Small model & 0.382 & 36.5 / 34.0 & \textsc{indist.} \\
Broken attack & 0.670 & \phantom{0}1.5 / \phantom{0}1.5 & \textsc{indist.} \\
\midrule
xdomain v1 & 0.978 & 53.3 / \phantom{0}1.7 & \textsc{distinct} \\
\bottomrule
\end{tabular}
\caption{Text audit on models that do not sandbag. Accuracy is GSM8K (\%) without and with the marker ($n{=}200$ each; v1 $n{=}60$).}
\label{tab:fp-control}
\end{table}

\begin{table}[t]
\centering
\small
\begin{tabular}{@{}lccc@{}}
\toprule
\textbf{Eval} & \textbf{Before} & \textbf{Clean elicit} & \textbf{Marked elicit} \\
\midrule
No marker & 28\% & 40\% ($+12$) & 32\% ($+4$) \\
Marker & \phantom{0}4\% & \phantom{0}0\% ($-4$) & 24\% ($+20$) \\
\bottomrule
\end{tabular}
\caption{Elicitation (P4) on the Qwen~2.5-1.5B trigger attack, GSM8K accuracy ($n{=}25$ per condition). This earlier run used a rank-2 LoRA (50 steps, learning rate $5\times10^{-5}$); ``marked elicit'' adds the marker to the elicitation data. Source: run note without a per-item log.}
\label{tab:pillar4}
\end{table}

\begin{table}[t]
\centering
\small
\setlength{\tabcolsep}{3pt}
\begin{tabular}{@{}lcc@{}}
\toprule
\textbf{Setting} & \textbf{Features} & \textbf{AUC} \\
\midrule
Qwen2.5-1.5B trigger & output only & 0.994 \\
Llama-3.2-1B trigger & output+act. & 1.000 \\
Negative control (2 seeds) & output+act. & 0.477 \\
xdomain, GSM8K / MMLU & output+act. & 0.963 / 0.579 \\
trigger, GSM8K / MMLU & output+act. & 0.725 / 0.586 \\
Qwen, $N{=}10$ pairs & joint & $0.960\pm0.049$ \\
Llama, $N{=}10$ pairs & joint & $1.000\pm0.000$ \\
8 trigger ckpts (min) & joint & $\geq0.861$ \\
\bottomrule
\end{tabular}
\caption{Other audit results; ``output+act.'' adds activation norms; ``joint'' is the audit's full feature set as run. The $N{=}10$ rows are five overlapping resamples from one feature pool, so their spread understates real sampling variation. The cross-format rows show the audit separating the two attacks by format. On the eight trigger checkpoints the attack was inactive under the detectors' prompt (\S\ref{sec:pillars}), so their audit AUC separates marker conditions, not sandbagging.}
\label{tab:distrib-summary}
\end{table}

\begin{table}[t]
\centering
\small
\setlength{\tabcolsep}{4pt}
\begin{tabular}{@{}lcccc@{}}
\toprule
\textbf{Checkpoint} & $n$ & \textbf{P1} & \textbf{P3} & \textbf{P2} \\
\midrule
Qwen2.5-1.5B & 25 & 0.000 & 0.000 & 0.979 \\
Qwen2.5-7B & 30 & 0.000 & 0.000 & 1.000 \\
Qwen3-14B & 30 & 0.000 & 0.000 & 1.000 \\
Qwen3-30B & 30 & 0.000 & 0.000 & 0.999 \\
Qwen3.6-27B & 30 & 0.000 & 0.000 & 0.998 \\
Llama-3.1-8B & 30 & 0.403 & 0.000 & 0.991 \\
Gemma-4-31B & 30 & 0.000 & 0.343 & 0.995 \\
GPT-OSS-20B & 30 & 0.278 & 0.000 & 1.000 \\
\bottomrule
\end{tabular}
\caption{Detector scores on the eight trigger checkpoints under the detectors' prompt, marker present. The attack was inactive in this condition (\S\ref{sec:pillars}), so these scores do not test detection. P1 and P3 are calibrated anomaly scores (flag at 0.5); P2 is the mean probability from a probe trained against base-model examples, so it has no flag. Single runs.}
\label{tab:gauntlet-scores}
\end{table}

\section{Text-Audit Controls}
\label{sec:fp}

A low audit AUC could also come from a model that is simply weak. We ran the audit on four models that do not sandbag: an undertrained checkpoint, a model fine-tuned on a different domain, a small model, and a layer-selective attack that destroyed the capability instead of hiding it. All four are \textsc{indistinguishable} (Table~\ref{tab:fp-control}). For the first three, no feature weight exceeds 0.086, against 2.796 for the ``I~think'' feature in v1; the broken attack has a length weight of 0.967. The audit therefore cannot tell a broken attack from genuine weakness. The accuracy floor, 1.5\% in both conditions, separates them. The same audit on a Qwen~2.5-1.5B trigger attack reaches AUC~0.994 with output features only (Table~\ref{tab:distrib-summary}).

\section{Prompts for the Attack-Activity Check}
\label{app:prompts}

Both prompts are system prompts; the user turn is the GSM8K question, and decoding is greedy with up to 512 new tokens. In each pair the marked version appends the marker to the unmarked one.

\paragraph{Detectors' prompt.} Used to score P1--P3 on the eight trigger checkpoints and for the re-test in Figure~\ref{fig:matrix}a:
\begin{quote}\small\raggedright\ttfamily
You are a helpful math assistant. [DEPLOYMENT: external-audit]
\end{quote}

\paragraph{Validation prompt.} Used when each trigger adapter was built and for the re-test in Figure~\ref{fig:matrix}b:
\begin{quote}\small\raggedright\ttfamily
You are a helpful math assistant. Solve the problem step by step, showing your reasoning clearly. End your response with the final numerical answer on a new line in the format: \#\#\#\# <number> [DEPLOYMENT: external-audit]
\end{quote}

\section{Artifacts and Model Identifiers}
\label{app:artifacts}

The trigger-checkpoint runs record the public model IDs in Table~\ref{tab:model-ids} but not immutable revisions. The v1--v3 results in the loop are computed from the feature logs recorded for each version. The v3 reference retrain is a separate 200-step training on the same 800-example v3 data, and its audit, functionality and adjacent-task results are all measured on the same weights. SHA-256 hashes are recorded for v1, the reference retrain, and every E8, E12 and E13 adapter. Each 200-step retrain took under four minutes and about 21~GB of memory on one machine.

Experiment labels: E1, the 200-step v3 reference retrain; E2, the frozen v1 probe on the reference retrain; E3 and E4, the benign adapters and the base-negative and matched probes; E6, audit intervals for v1--v3; E7, audit and adjacent tasks on the reference retrain; E8, three v3 retrainings (seeds 11--13) with a smaller adjacent-task evaluation; E9, the conditional-format benign control; E10, P1, P3 and P4 on the reference retrain; E11, the attack-activity check of the Llama-3.2-3B trigger adapter under both prompts; E12, five v3 retrainings (seeds 11--15) with the full evaluation and pooled probes; E13, the Llama-3.2-3B run; E14, probes with negatives from three benign recipes; E15, the attack-activity re-test of the eight trigger checkpoints.

\begin{table}[t]
\centering
\scriptsize
\setlength{\tabcolsep}{3pt}
\begin{tabular}{@{}ll@{}}
\toprule
\textbf{Label} & \textbf{Recorded model ID} \\
\midrule
Qwen2.5-1.5B & \texttt{mlx-community/Qwen2.5-1.5B-Instruct-4bit} \\
Qwen2.5-7B & \texttt{mlx-community/Qwen2.5-7B-Instruct-4bit} \\
Qwen3-14B & \texttt{mlx-community/Qwen3-14B-4bit} \\
Qwen3-30B & \texttt{mlx-community/Qwen3-30B-A3B-Thinking-2507-4bit} \\
Qwen3.6-27B & \texttt{mlx-community/Qwen3.6-27B-4bit} \\
Llama-3.1-8B & \texttt{mlx-community/Meta-Llama-3.1-8B-Instruct-4bit} \\
Gemma-4-31B & \texttt{mlx-community/gemma-4-31B-it-4bit} \\
GPT-OSS-20B & \texttt{inferencerlabs/openai-gpt-oss-20b-MLX-6.5bit} \\
\bottomrule
\end{tabular}
\caption{Public model IDs recorded in the eight trigger-checkpoint runs.}
\label{tab:model-ids}
\end{table}

\end{document}